\documentstyle[prl,aps,amsmath,latexsym,amssymb]{revtex}
\def \d{\mbox{d}}
\def \D{\mbox{D}}

\def \pd{\partial}

\def \iner{\rfloor}
\def \hodge{{}^\star}
\def \stimes{\times\!\!\!\!\!\!\supset}

\def \tl#1{\overset{\kern 1pt\circ}{#1}}
\def \TL#1{\overset{\kern -3pt \circ}{#1}}
\def \TLL#1{\overset{\kern -7pt \circ}{#1}}

\begin{document}

\title{Dislocation theory as a 3-dimensional translation gauge theory\footnote{Dedicated to 
	Professor Ekkehart Kr{\"o}ner, Stuttgart, on the occasion of his 80th birthday.}}
\author{ Markus Lazar\footnote{Present address: Institute for Theoretical Physics,
     University of Leipzig, Augustusplatz 10, D--04109 Leipzig, 
     Germany, E-mail: lazar@itp.uni-leipzig.de.}}
\address{Max-Planck-Institut f{\"u}r Metallforschung\\ 
     Institut f{\"u}r Physik\\
     D--70569 Stuttgart, Germany}

\date{
      \today}
\maketitle
\begin{abstract}
We consider the static elastoplastic theory of dislocations in an elastoplastic 
material. We use a Yang-Mills type Lagrangian (the teleparallel equivalent of
Hilbert-Einstein Lagrangian) and some Lagrangians with anisotropic constitutive
laws. The translational part of the generalized affine connection is utilized to 
describe the theory of elastoplasticity in the framework of a translation gauge 
theory. We obtain a system of Yang-Mills field equations which express the 
balance of force and moment.
\\

\end{abstract}
\vspace*{5mm}
\noindent
{\it to appear in Annalen der Physik (2000)}
\section{Introduction}
The field theory of defects in crystals has an old history.
First, Kondo~\cite{kondo52} and Bilby et al.~\cite{bilby55}
described independently a dislocation theory in the language of differential geometry.
They proved the equivalence of dislocation density to Cartan's torsion.
Kr{\"o}ner and Seeger~\cite{kroener59,kroener60} completed this theory 
to a non-linear theory of elasticity with dislocations and internal stress. 
Recent developments of this theory are given in~\cite{kroener90,kroener95}.

The first step towards a gauge theory of defects had been taken by 
Turski~\cite{turski66}. He derived the equilibrium equations by means of a
variational principle.

Edelen et al.~\cite{edelen83,edelen88} developed a theory of
dislocations and disclinations as a gauge theory of the Euclidean group 
${SO(3)\stimes T(3)}$ and a pure dislocation theory as a translational or 
$T(3)$-gauge theory. Unfortunately, they did not distinguish between disclinations 
in solid and in liquid crystals. Moreover, they did not refer to the gauge 
theory of the Poincar{\'e} group ${SO(1,3)\stimes T(4)}$ which was very well 
developed in gravity~\cite{hehl76}. 
Their theory is claimed to describe the dynamics of defects. 
However, they have ignored the following effects:
Dislocations can move in two different modes, called glide and climb. 
If dislocations climb, they interchange with point defects such as vacancies
and/or interstitials.
Therefore, it is necessary to include the point defects into a dynamical 
theory of dislocations.
Furthermore, the motion of dislocation is highly dissipative 
(friction and radiation damping). 

Other proposals for gauge theories of dislocations were put forward by 
Gairola~\cite{gairola81,gairola93}, Kleinert\cite{kleinert83,kleinert89}, 
Trz\c{e}sowski~\cite{trzesowski93}, and, quite recently, by Malyshev~\cite{malyshev99}.

The classical theory of dislocations deals with the incompatibility
and the equilibrium conditions~(see, e.g.,~\cite{seeger55}). 
The drawback of this theory is that a field equation is missing. Therefore, 
one has to take as ansatz for the dislocation density a singular 
$\delta$-function. The price to be paid is that the elastic energy is singular. 
Thus the information about the dislocation core is lost.

The aim of this paper is to develop a $T(3)$-gauge theory of dislocations 
similar to gravity as $T(4)$-gauge theory. These gauge theories are included
in the metric-affine gauge theory~(MAG) as given by Hehl et al.~\cite{hehl95},
where the metric, the coframe, and the connection are independent field
variables. We use the tool of MAG in order to derive the gauge theory of 
dislocations in an elastoplastic material as field theory of elastoplasticity.

We obtain a system of static field equations which determine both, the elastic 
and the plastic fields. Additionally, the equations contain the equilibrium 
condition and the Bianchi identity for the torsion (or dislocation density). 
As shown later, our field equations differ from the field equations proposed 
by Edelen et al.~\cite{edelen83,edelen88}. One reason is that we have used 
a different and more physically motivated gauge Lagrangian. For compact notation, 
we will use the calculus of exterior differential forms.

\section{Geometrical framework}
In order to gauge the translation group, it is convenient to start with
the affine group. 
Therefore, we consider the affine group
in three dimensions $A(3)={GL(3)\stimes T(3)}$, where the symbol
$\stimes$ denotes the semidirect product.\\
The mathematical domain of gauge theories is the theory of fibre 
bundles~\cite{trautman80,mielke}; for rigorous definitions of fibre bundles and 
connections, see~\cite{koba1,trautman84}.
In the case of the affine group, the corresponding fibre bundle is the
bundle of affine frames $AM$. 
The connection 1-form $\widetilde{\widetilde{\omega}}$ on $AM$ is called 
{\it generalized affine connection}. It is an ${\frak a}(3)$-valued 1-form
of type ad, where ${\frak a}(3)$ is the Lie algebra of $A(3)$. 
The corresponding 2-form $\widetilde{\widetilde{\Omega}}$ is the curvature 2-form
on $AM$. It is given by Cartan's structure equation
\begin{align}
\widetilde{\widetilde{\Omega}}:=\widetilde{\widetilde{\D}}\widetilde{\widetilde{\omega}} 
            \equiv\d\widetilde{\widetilde{\omega}}+\widetilde{\widetilde{\omega}}\wedge\widetilde{\widetilde{\omega}} .
\end{align}
The symbol $\wedge$ denotes the exterior product, $\d$ is the exterior derivative 
and $\D$ the gauge covariant exterior derivative.

Using the subbundle map $\gamma: LM\mapsto AM$, 
$\widetilde{\widetilde{\omega}}$ can be pulled back to the bundle of linear frames $LM$,
\begin{align}\label{Kon-aff}
\gamma^{*}\,\widetilde{\widetilde{\omega}}=\omega+\phi=\omega^{ab}L_{ab}+\phi^a P_a,
\end{align}
where  $\omega$ is a ${\frak gl}(3)$-valued 1-form on $LM$, 
the {\it linear connection}, and $\phi$ is an ${\Bbb R}^{3}$-valued tensorial 
1-form on $LM$, the {\it translational connection}. Here $P_a$ are the generators 
of translations ${\Bbb R}^{3}$ and the generators $L_{ab}$ span the Lie algebra 
${\frak gl}(3)$.
The pull back of $\widetilde{\widetilde{\Omega}}$ is given by
\begin{align}\label{affKruem}
\gamma^{*}\,\widetilde{\widetilde{\Omega}}=\Omega+\Phi
=\Omega^{ab}L_{ab}+\Phi^a P_a,
\end{align}
where
\begin{align}
\Omega:=\D\omega=\d\omega+\omega\wedge\omega
\end{align}
is the curvature of the linear connection $\omega$. The {\it translational
curvature} $\Phi$ reads
\begin{align}
\label{trans-curv}
\Phi:=\D\phi=\d\phi+\omega\wedge\phi.
\end{align}
In particular, the translational connection can be decomposed into
the soldering 1-form $\vartheta$ and an additional 1-form $\chi$,
\begin{align}
\phi=\vartheta+\chi .
\end{align}
In order to ensure the tensorial transformation behaviour of 
$\vartheta=\vartheta^a P_a$, which is an ${\Bbb R}^{3}$-valued tensorial 1-form of 
type id, it is convenient to follow Trautman~\cite{trautman73}. We introduce the 
vector-valued 0-form $\xi\equiv\xi^a P_a$ and identify $\chi=-\D\xi$.
With these identifications, the soldering-form turns out to be
\begin{align}
\label{soldering-form}
\vartheta=\widetilde{\widetilde{\D}}\xi\equiv\D\xi+\phi.
\end{align}
It transforms as a vector-valued 1-form.
With (\ref{trans-curv}) and (\ref{soldering-form}), the torsion 2-form,
in the framework of the affine group, reads 
\begin{align}
T\equiv T^a P_a:=\D\vartheta=\big(\Phi^a+\Omega^a_{\ b}\xi^b\big)P_a.
\end{align}

Since the translation group is a subgroup of $A(3)$, we obtain, 
with the choice $\omega\equiv 0$, the translational part of the 
{\it generalized affine connection} as 
\begin{align}
\phi=\vartheta-\d\xi .
\end{align} 
Some remarks on the generalized affine connection in relation to
dislocations have been made by Mistura~\cite{mistura90}.

\section{Physical identifications and classical elasticity}
The soldering form is given with respect to the translational part of the 
generalized affine connection $\phi^a$ by
\begin{align}
\label{key-rel}
\vartheta^a=\d \xi^a+\phi^a.
\end{align}
Eq.~(\ref{key-rel}) is the key formula for the $T(3)$-gauge theory.
It is valid in a Weitzenb{\"o}ck space ($\Omega^{a}_{\ b}= 0$) under the 
gauge condition $\omega^{a}_{\ b}\equiv 0$. 
The coupling in eq.~(\ref{key-rel}) between the translational gauge potential 
$\phi^a$ and the vector field $\xi^a$ is characteristic for the 
translation gauge theory as a special case of the metric--affine gauge theory. 
The reason lies in the nature of the affine group itself.

At first, in agreement with  Edelen~\cite{edelen83}, we identify the 
translational connection or the translational gauge potential $\phi^a$ as the 
gauge potential of the dislocations.
This identification is justified by the following ideas.
A gauge potential is a 1-form. If we form the corresponding 2-form,
there results the gauge field strength. 
In our case of a Weitzenb{\"o}ck space, in the gauge chosen, the torsion 
2-form is equal to the object of anholonomity. Therefore, one obtains
\begin{align}
T^a=\d\vartheta^a\equiv\d\phi^a
=\frac{1}{2} T^a_{\ ij}\,\d x^i\wedge\d x^j,
\end{align}
which can be identified with the dislocation density. 
One recovers the conventional dislocation density tensor $\alpha^a_{\ i}$ from
$T^a$ by means of $\alpha^a_{\ i}=\frac{1}{2}\epsilon_i^{\ jk} T^a_{\ jk}$.
Thus, the dislocation density or the torsion represents the gauge field strength 
of the translation group. In the framework of $T(3)$-gauge theory 
($\omega^{a}_{\ b}\equiv 0$), the torsion 2-form satisfies the Bianchi identity
\begin{align}
\label{1.Bianchi}
\d T^a=0 .
\end{align}

The indices $a,b,c,\ldots=1,2,3$ denote the (anholonomic) material or the  
final coordinates and $i,j,k,\ldots=1,2,3$ the (holonomic) Cartesian 
coordinates of the reference system (defect-free or ideal reference 
system)\footnote{A field of coframes $\vartheta$ is {\it holonomic} if
$\d\vartheta=0$ and {\it anholonomic} if $\d\vartheta\neq 0$\cite{schouten}.}.
Now, we identify $\xi^a$ with the diffeomorphisms of the material space into
the Euclidean space in which the crystal is embedded. We introduce
the displacement field $u^a$ and write $\xi^a$ in terms of it as 
$\xi^a=\delta^a_i x^i+u^a$.
With this identification, the soldering form is specified
\begin{align}
\label{el-distorsion}
\vartheta^a=B^a_{\ i}\,\d x^i
=\big(\delta^a_i+\pd_i u^a+\phi^a_{\ i}\big)\d x^i.
\end{align}
The following consideration justifies the last identification:
In a material with compatible distortion $\phi^a=0$, the soldering form is
$\vartheta^a=\d\xi^a$.
The metric of the final state is the Cauchy-Green strain tensor $G$ which, 
in our case (teleparallelism), is given by
\begin{align}
G=\delta_{ab}\,\vartheta^a\otimes\vartheta^b
=B^a_{\ i}B_{aj}\,\d x^i\otimes\d x^j
=g_{ij}(x)\d x^i\otimes\d x^j,\qquad 
\delta_{ab}=\text{diag}(+++).
\end{align}
The engineering strain tensor is
\begin{align}
2E=G-1=(g_{ij}-\delta_{ij})\d x^i\otimes\d x^j.
\end{align}
It is obvious that the soldering 1-form corresponds to the distortion
1-form. 
If the distortion is compatible, the distortion is given as a deformation 
gradient of a vector field. This vector field is $\xi^a$.

In the general case of elastoplasticity, eqs.~(\ref{key-rel}) and 
(\ref{el-distorsion}) describe an incompatible distortion with the 
distortion 1-form ${\boldsymbol \beta}\equiv{\boldsymbol\vartheta}$. 
The cause of plasticity are defects and the material gives rise to a specific 
elastic response. 
The plastic distortion is given by the dislocation gauge potential 
${\boldsymbol \beta}^{\rm pl}\equiv{-\boldsymbol\phi}$.
Finally, the total distortion ${\boldsymbol \beta}^{\rm T}$ contains elastic 
and plastic contributions according to
\begin{align}
{\boldsymbol \beta}^{\rm T}= {\boldsymbol \beta}^{\rm pl}+{\boldsymbol \beta}.
\end{align}
Since it is compatible, we have
\begin{align}
\d{\boldsymbol \beta}^{\rm T}=0,\qquad\d{\boldsymbol \beta}^{\rm pl}=-\d{\boldsymbol \beta}.
\end{align}
Thus the physical space (crystal) is determined by the two 
fields $\d\xi^a$ and $\phi^a$.

Let us now discuss the concepts of {\it internal} and {\it external observers}
as introduced by Kr{\" o}ner~\cite{kroener81}.
The internal observer lives in the crystal and uses $\vartheta^a$ as 
coframe. Consequently, he can detect defects due to $\d\vartheta^a=T^a$.
But he misses the information about the holonomic coordinate system
and is unable to detect compatible deformations.
The external observer lives in the external space in which the crystal 
is embedded. He has more information available than the internal observer 
because he knows the holonomic coordinate system and is able to detect
compatible deformations.
The external observer measures the torsion in his coordinate system as
$T^i=B_a^{\ i} T^a$.

In gravity, the physical meaning of $\xi^a$ is not well-understood~\cite{hehl95,hehl98}.
It is obvious that we are internal observers in our universe and cannot
detect $\xi^a$ or $\d\xi^a$. The internal observer in gravity would use the 
{\it Cartan} or {\it affine connection} 
($\vartheta^a,\omega^a_{\ b}$) instead of the generalized affine connection.
Does, perhaps, this idea also lead to an understanding of $\xi^a$ in gravity?

In order to simplify the formulas, we assume a linear constitutive law 
(stress-strain relation). But our general considerations are not restricted to 
this assumption. The elastic (anisotropic) behaviour of the material can be
described by the elasticity tensor~\cite{marsden}
\begin{align}
C=C^{ijkl}\,\pd_i\otimes\pd_j\otimes\pd_k\otimes\pd_l .
\end{align}
$C$ is a contravariant tensor of fourth rank with the symmetries
\begin{align}
C^{ijkl}=C^{jikl}=C^{ijlk}=C^{klij} .
\end{align}
The elastic energy (potential energy) contains the constitutive law and 
is given by
\begin{align}
W= \frac{1}{2} C:(E\otimes E)
 =\frac{1}{2} C^{ijkl}E_{ij}E_{kl},
\end{align}
where the symbol $:$ denotes double contraction.
Since we consider a static and not a dynamical theory, the elastic 
Lagrangian is given by means of the potential energy
\begin{align}
{\cal L}_{\mathrm{el}}= -\,\hodge W=-W\d v_{\rm E},
\end{align}
where the volume 3-form is defined by
\begin{align}
\d v_{\rm E}:=\frac{1}{3!}\eta_{abc}\,\vartheta^a\wedge\vartheta^b\wedge\vartheta^c ,
\end{align}
with $\eta_{abc}:= \sqrt{|B|}\epsilon_{abc}$, and $\epsilon_{abc}$ as 
the Levi-Civita symbol. Furthermore, $\eta_a:=e_a\iner\d v_{\rm E}$, 
$\eta_{ab}:=e_a\iner e_b\iner\d v_{\rm E}$, and 
$\eta_{abc}:=e_a\iner e_b\iner e_c\iner\d v_{\rm E}$, where
$\iner$ denotes the interior product with
\begin{align}
e_a\iner\vartheta^b=B_a^{\ i}B^b_{\ i}=\delta_a^b,\qquad e_a=B_a^{\ i}\pd_i.
\end{align}
The symbol $\hodge$ denotes the Hodge star operator which, in three dimensions, 
defines the dual $(3-p)$-form of a given $p$-form.

\section{Gauge theory of dislocations}
Let us now derive the gauge Lagrangian of the translation group $T(3)$ in
analogy to gravity~\cite{hehl95,Erice95,MAG}.
We make the most general Yang-Mills ansatz which is quadratic 
in the corresponding field strength
$T^a=\d\vartheta^a$,
\begin{align}
V_\|=-\frac{1}{2}T^a\wedge H_a .
\end{align}
The simplest choice is $H_a=\frac{1}{\ell}\,\hodge T_a$ which was used by Edelen 
for a dislocation gauge theory. 
It is well-known that, in gravity, this Lagrangian does not yield Einstein's 
theory~\cite{thirring}.    
Recently, Malyshev~\cite{malyshev99} discussed the gauge Lagrangian used 
by Edelen~\cite{edelen83,edelen88}. He showed that it does not lead to the correct 
solutions for edge dislocations within a linear approximation.
Accordingly, we will use the most general Lagrangian for an isotropic material 
with the 1-form
\begin{align}
\label{const_iso}
H_a=\frac{1}{\ell}\,\hodge\!\sum_{I=1}^{3}a_{I}\,^{(I)}T_a.
\end{align}
Here $\ell$ is the coupling constant of the theory.
It is obvious that eq.~(\ref{const_iso}) is a constitutive law for an 
isotropic material.
This Lagrangian has to be invariant under local $SO(3)$-transformations
in order to obtain the teleparallel version of the Hilbert-Einstein Lagrangian,
since the Einstein theory of gravity can be viewed as the gauge theory
of the translation group in four dimension. In this framework, the vierbein fields 
are the translational gauge potentials and the field strength is given in terms
of the anholonomity~\cite{cho}.

The gauge Lagrangian~(\ref{const_iso}) was used by Katanaev and Volovich~\cite{katanaev92} 
to describe dislocations, but without combining it with an elastic 
Lagrangian. Thus their field equations do not really describe
dislocations in an elastic material.

The condition of local $SO(3)$-invariance  yields the following three 
parameters~(see, e.g.,~\cite{katanaev92},\cite{hehl95}), 
\begin{align}
a_1=-1,\qquad a_2=2,\qquad a_3=\frac{1}{2}\,.
\end{align}
Consequently
\begin{align}
\label{const-tele}
H_a=\frac{1}{\ell}\,\hodge\big(- ^{(1)}\!T_a+2 \,^{(2)}T_a+\frac{1}{2}\,^{(3)}T_a\big).
\end{align}
The three pieces $^{(I)}T_a$ are irreducible with respect to  $SO(3)$.
Thus the torsion or dislocation density reads
$T^a=\,^{(1)}T^a+\,^{(2)}T^a+\,^{(3)}T^a$, with the number of
independent components $9=5\oplus 3\oplus 1$, where (for notations see \cite{hehl95})
\begin{alignat}{2}
^{(1)}T^a&:=T^a-\,^{(2)}T^a-\,^{(3)}T^a
&&\qquad \text{(tentor)},\\
^{(2)}T^a&:=\frac{1}{2}\vartheta^a\wedge\big(e_b\iner T^b\big)
&&\qquad\text{(trator)},\\
^{(3)}T_a&:=\frac{1}{3}e_a\iner\big(\vartheta^b\wedge T_b\big)
&&\qquad\text{(axitor).}
\end{alignat}
In order to obtain these three irreducible pieces, one can use the standard 
method of Young tableaux.
In the field theory of dislocations, the axitor describes three vertical bands 
of screw dislocations~\cite{hehl65}. 

With the identity ($R^{ab}$=Riemann-Cartan curvature 2-form,
$\widetilde{R}^{ab}$=Riemann-Christoffel curvature 2-form)
\begin{align}
\widetilde{R}^{ab}\wedge\eta_{ab}=R^{ab}\wedge\eta_{ab}
+T^a\wedge \hodge\big(-^{(1)}\!T_a+2 \,^{(2)}T_a+\frac{1}{2}\,^{(3)}T_a\big)
+2\d\big(\vartheta^a\wedge \hodge T_a\big),
\end{align}
in the case of teleparallelism  $R^{ab}=0$ 
and dropping the surface term, we find
the equivalence to the Hilbert-Einstein Lagrangian $V_{\rm GR}$,
\begin{align}
\label{Einstein-Hilbert}
V_\|=V_{\rm GR}=-\frac{1}{2\ell}\widetilde{R}^{ab}\wedge\eta_{ab}
\equiv  \frac{1}{2\ell}\,\hodge\widetilde{R}\,.
\end{align}
Here $\widetilde{R}:=e_a\iner e_b\iner \widetilde{R}^{ab}$.
Finally, the total Lagrangian reads
\begin{align}
{\cal L}=V_\|+{\cal L}_{\rm el}.
\end{align}

Thus, for the screw dislocations in an isotropic material, 
one needs only one material constant. 
Accordingly, in this specific case, the ansatz of Edelen et al.~\cite{edelen83} 
coincides with the more general ansatz~(\ref{const_iso}). Moreover, 
in the framework of Edelen's dislocation theory, 
Osipov\cite{osipov} showed that the second order corrections to the stress field 
of a {\it screw} dislocation far from the core are in good agreement with the 
formulas given by Kr{\"o}ner and Seeger~\cite{kroener59}. 

Let us now turn to the anisotropic case.
The linear constitutive law for an anisotropic material reads
\begin{align}
\label{const_ani_lin}
H_a=\frac{1}{2}\,\hodge\Big(\kappa_{aij}^{\ \ \ \, bkl}\, T_{bkl}\,\d x^i\wedge\d x^j\Big).
\end{align}
A non-linear constitutive law, in analogy to non-linear (Born-Infeld type) 
electrodynamics, is given by
\begin{align}
\label{const_ani_nonlin}
H_a=\frac{1}{2}\,\hodge
\Big(\kappa_{aij}^{\ \ \ \, bkl}\, T_{bkl}\,\d x^i\wedge\d x^j +
\tilde{\kappa}_{aij}^{\ \ \ \, bklcmn}\, T_{bkl}\,T_{cmn}\,\d x^i\wedge\d x^j
+\ldots\Big),
\end{align}
where $\kappa_{aij}^{\ \ \ \, bkl}$ and 
$\tilde{\kappa}_{aij}^{\ \ \ \, bklcmn}$ are the constitutive functions.

The elastic stress tensors, or the corresponding stress forms, are the
currents of the elastic field.
Therefore, the stress 2-form is defined by 
\begin{align}
\Sigma_a:=\frac{\delta {\cal L}_{\rm el}}{\delta \vartheta^a}.
\end{align}
In local components, this stress 2-form reads,
\begin{align}
\Sigma_a=\frac{1}{2}\Big(\sigma^{kl}B_{ak}\eta_{lmn}
	-\frac{1}{2}\sigma^{kl}E_{kl}\eta_{amn}\Big)\d x^m\wedge\d x^n,
\end{align} 
with $\sigma^{kl}=C^{ijkl}E_{ij}$. The first Piola-Kirchhoff stress tensor 
is included  in the 1-form dual to the stress 2-form 
\begin{align}
\hodge\Sigma_a=t_{al}\,\d x^l
=\Big(\sigma^{k}_{\ l}B_{ak}-\frac{1}{2}\sigma^{ij}E_{ij}B_{al}\Big)\d x^l,
\end{align} 
where $t_{al}$ is the first Piola-Kirchhoff stress tensor. 
The second Piola-Kirchhoff stress tensor, derived from the first Piola-Kirchhoff
stress tensor, is given by
\begin{align}
t_{ac}=B_c^{\ l}\, t_{al}
=\sigma_{ac}-\frac{1}{2}\delta_{ac}\,\sigma^{ij}E_{ij}.
\end{align}
Here $t_{ac}$ corresponds to the Maxwell tensor of elasticity
for the incompatible case.
The symmetric Cauchy stress tensor arises as
\begin{align}
t_{ij}=B^a_{\ i}\, t_{aj}
=\sigma_{ij}-\frac{1}{2}g_{ij}\,\sigma^{kl}E_{kl}.
\end{align}
If dislocations are present, the stress forms are the elastic responses of the 
body to the dislocations.

The gauge field momentum 1-form (nowadays called excitation) is defined by
\begin{align}
H_a:=-\frac{\pd V_\|}{\pd T^a},
\end{align}
i.e., it is the specific ``response'' quantity of the gauge Lagrangian $V_\|$ 
(and not of the elastic Lagrangian ${\cal L}_{\rm el}$) to $T^a$.
It has the dimension of a moment stress. 
Thus, we interpret $H_a$ as a moment or couple stress originating in 
the dislocation core. The moment stress 1-form, in local components 
for an anisotropic linear material, reads
\begin{align}
H_a=H_{an}\d x^n
   =\kappa_{a}^{\ \,ijbkl}\, T_{bkl}\, \eta_{ijn}\,\d x^n.
\end{align}

On the other hand, the translation gauge current or the stress 2-form of the 
gauge field,
\begin{align}
h_a:=\frac{\pd V_\|}{\pd \vartheta^a}
=e_a\iner V_\|+\big(e_a\iner T^b\big)\wedge H_b,
\end{align}
explicitly reads
\begin{align}
h_a=\frac{1}{2}\Big[\big(e_a\iner T^b\big)\wedge H_b
-T^b \big(e_a\iner H_b)\Big].
\end{align}
Whereas $\Sigma_a$ is the stress 2-form of the elastic material,  
$h_a$ is the response stress 2-form of the dislocations or the plastic stress 
2-form. Accordingly, $h_a$ is the internal stress caused by dislocations.

The field equations are derived by varying the total Lagrangian with respect 
to the elastic field $\xi^a$ and the gauge potential $\phi^a$:
\begin{align}
\frac{\delta{\cal L}}{\delta \xi^a}
&\equiv
\d\Sigma_a+\d h_a=0\label{FG1},\\
\frac{\delta{\cal L}}{\delta \phi^a}
&\equiv
\d H_a - h_a=\Sigma_a .
\label{FG2}
\end{align}
Eq.~(\ref{FG2}) is the Yang-Mills type gauge field equation 
in the theory of elastoplasticity.
Both stress 2-forms, $\Sigma_a$ and $h_a$, are sources in eq.~(\ref{FG2}).
Eq.~(\ref{FG1}) is a consequence of (\ref{FG2}) and the Poincar{\'e} lemma 
$\d \d=0$. 
The Euler-Lagrange equations can be interpreted as equilibrium equations. 
Eq.~(\ref{FG1}) describes the force and eq.~(\ref{FG2}) the moment equilibrium. 
The fields to be determined from the field equations are $\xi^a$ or $u^a$
and $\phi^a$. 
Due to the  nonlinear geometrical character of elastoplasticity,
the field equations are represented by a coupled system of nonlinear partial 
differential equations.
In the framework of MAG, eq.~(\ref{FG1}) is the matter field and
eq.~(\ref{FG2}) the first gauge field equation (see~\cite{hehl95}).

The force density 3-forms are defined as follows:
The Peach-Koehler force~\cite{peach} (elastic force acting on a dislocation), 
which is analogously defined as the Lorentz force in Maxwell's theory,
reads
\begin{align}
\label{PK-force}
f^{\rm el}_a:=\d\Sigma_a=\big(e_a\iner T^b\big)\wedge\Sigma_b.
\end{align}
The response force to dislocations is
\begin{align}
\label{response-force}
 f^{\rm g}_a:=\d h_a=\big(e_a\iner T^b\big)\wedge h_b.
\end{align}
Eq.~(\ref{FG1}) describes the force equilibrium between the 
Peach-Koehler force and the dislocation-response force
\begin{align}
\label{ges-forces}
0=f_a=f^{\rm g}_a+f^{\rm el}_a=\d h_a+\d\Sigma_a,
\end{align}
or is, equivalently, interpreted as the equilibrium condition of an elastic body 
containing dislocations.
Both these forces are configurational forces.
It is to be emphasized that (\ref{PK-force}) and (\ref{response-force}), 
in the framework of field theory, are the 
first Noether identities for the elastic and the gauge fields.

For vanishing dislocation density ($\phi^a=0$ and $T^a=0$), 
the field equations reduce to the equilibrium condition of the 
classical elasticity theory under zero external forces
\begin{align}
\d\Sigma_a=0\quad\Rightarrow\quad \pd^i t_{ai}=0 .
\end{align}

\section{Relations to other descriptions of dislocation theory}
In an alternative description of teleparallelism one replaces 
$T^a\equiv\d\vartheta^a$ by the Levi-Civita connection 
$\widetilde{\omega}^a_{\ b}$ of the metric (Cauchy-Green tensor) 
$G=\delta_{ab}\vartheta^a\otimes\vartheta^b$.
One applies Cartan's first structure equation
\begin{align}
\label{1CE}
\d\vartheta^a=-\widetilde{\omega}^a_{\ b}\wedge\vartheta^b,
\end{align}
which yields
\begin{align}
\widetilde{\omega}_{ab}=
\frac{1}{2}\big(
e_a\iner T_b-e_b\iner T_a-\big(e_a\iner e_b\iner T_c\big)\wedge\vartheta^c
\big).
\end{align} 
The corresponding Riemannian curvature 2-form reads
\begin{align}
\label{Riem-curv}
\widetilde{R}_{ab}=\d\widetilde{\omega}_{ab}+\widetilde{\omega}_{ac}\wedge\widetilde{\omega}^c_{\ b}.
\end{align}
Eventually, we get the corresponding field equation by substituting eq.~(\ref{Riem-curv})
into the Hilbert-Einstein Lagrangian $V_{\rm GR}$ of eq.~(\ref{Einstein-Hilbert}). 
After variation, one recovers the Einstein type field equation
\begin{align}
\label{Einstein-eq}
\frac{1}{2}\eta_{abc}\,\widetilde{R}^{bc}=\ell\,\Sigma_a
\end{align}
used by Malyshev~\cite{malyshev99} for dislocation theory.
A disadvantage of this description is that the original structures are blurred
and that the field equation does not have the Yang-Mills form. 

Now, we can completely translate our formulas of the gauge theoretical description 
of dislocation theory into the coordinate system of the 
external observer (holonomic coordinates).
The linear connection of the external observer is defined as pure gauge,
\begin{align}
\omega^i_{\ j}=B_a^{\ i}\d B^a_{\ j}=B_a^{\ i}\,\pd_k B^a_{\ j}\,\d x^k.
\end{align}
In holonomic coordinates, the torsion is given by 
\begin{align}
\label{torsion-holonom}
T^i=B_a^{\ i}\, \pd_{[j} B^a_{\ k]}\,\d x^j\wedge\d x^k=\omega^i_{\ k}\wedge\d x^k.
\end{align}
Cartan's torsion tensor $T^i_{\ jk}$ is the antisymmetric
part\footnote{We will be using the notation $A_{[ij]}\equiv\frac{1}{2}(A_{ij}-A_{ji})$.}
of the components of the connection $\omega^i_{\ jk}$, 
namely $\frac{1}{2}T^i_{\ jk}\equiv \omega^i_{\ [jk]}$.
By means of the torsion~(\ref{torsion-holonom})
and the condition of vanishing nonmetricity,
\begin{align}
\D g_{ij}\equiv
\d g_{ij}-\omega^k_{\ i} g_{jk}-\omega^k_{\ j} g_{ik}=0.
\end{align}
The connection $\omega^i_{\ j}$ can be decomposed
into the Levi-Civita connection $\widetilde{\omega}^{i}_{\ j}$ and the
contortion $\tau^{i}_{\ j}$, which is a tensorial 1-form of type ad, 
\begin{equation}
\omega^{i}_{\ j}=\widetilde\omega^{i}_{\ j} -\tau^i_{\ j}.
\end{equation}
The local components of the Levi-Civita connection and of the
contortion read, respectively,
\begin{equation}
\widetilde\omega^i_{\ j}=\frac{1}{2}g^{il}\big(\pd_{j}g_{lk}+\pd_{k}g_{lj}-\pd_{l}g_{jk}\big)
\d x^k
\end{equation}
and 
\begin{equation}
\label{contor}
\tau^{i}_{\ j}=\frac{1}{2}\left(T^{i}_{\ jk}+T_{jk}^{\ \ i}-T_{k\ j}^{\ i}\right)\d x^k.
\end{equation}
We can resolve (\ref{contor}) with respect to the torsion,
\begin{equation}
\tau^{i}_{\ j}\wedge\d x^{j}=-\frac{1}{2}T^{i}_{\ jk}\d x^{j}\wedge\d x^{k}
=-T^{i}.
\end{equation}
The forces are
\begin{align}
\label{PK-force2}
f^{\rm el}_i=\D\Sigma_i\equiv\d\Sigma_i+\omega_i^{\ j}\wedge\Sigma_j
	     = T^j_{\ i}\wedge\Sigma_j
\end{align}
and
\begin{align}
\label{response-force2}
f^{\rm g}_i=\D h_i = T^j_{\ i}\wedge h_j.
\end{align}
In holonomic coordinates, the field equations read
\begin{align}
&\D\Sigma_i+\D h_i=0\qquad\text{(force equilibrium)},\\
&\D H_i=h_i+\Sigma_i\qquad\text{(moment equilibrium)}.
\end{align}
Using the condition of teleparallelism $R^{ij}=0$, 
in linear approximation we obtain Kr{\"o}ner's incompatibility equation 
\begin{align}
{\rm inc}\,E\equiv\nabla\times E\times\nabla=\eta.
\end{align}
The symmetric second rank tensor $\eta$ is called incompatibility
tensor; it encompasses the dislocation density.
With $H_i=0$, we recover a dislocation theory without moment stress which
is in agreement with the dislocation theory given by Kr{\"o}ner and 
Seeger~\cite{kroener59,kroener60}.

\section{Conclusion}
We have proposed a dislocation gauge theory in an elastoplastic material.
The basic equations~(\ref{1.Bianchi}), (\ref{FG1}), (\ref{FG2}) and 
(\ref{ges-forces}) of dislocation gauge theory can be
summarized in axiomatic way in analogy to the Maxwell theory (for an axiomatic
formulation of Maxwell's theory, see, e.g.,~\cite{electro1,electro2}).
As soon as the Lagrangian is specified, one can find the basic laws. 
As gauge Lagrangians we use the teleparallel one, which is equivalent to the 
Hilbert-Einstein Lagrangian, and some Lagrangians for anisotropic constitutive laws. 
For linear constitutive laws, the total Lagrangian has the following symbolic form
\begin{align}
{\cal L}\sim (\text{strain})^2+(\text{dislocation density})^2.
\end{align}
The {\it first law} expresses the force equilibrium
\begin{align}
\label{law1}
\d\Sigma_a^{\rm T}=0\qquad\text{with}\qquad\Sigma_a^{\rm T}=\Sigma_a+h_a .
\end{align}
A consequence of eq.~(\ref{law1}), in analogy to the inhomogeneous Maxwell equation,
is the inhomogeneous Yang-Mills equation
\begin{align}
\d H_a=\Sigma_a^{\rm T}\qquad\text{(moment equilibrium)}.
\end{align}
The definition of the elastoplastic force is the {\it second law}
\begin{align}
f_a=(e_a\iner T^b)\wedge\Sigma_b^{\rm T}.
\end{align}
The conservation law of dislocation density (homogeneous Yang-Mills equation
or Bianchi identity) is the {\it third law}
\begin{align}
\d T^a=0.
\end{align}

The constitutive laws ${\sigma}\sim E$ and $H\sim T$, which are 
the physical input from experimental data, are the {\it fourth law}.

We have compared our proposal with Kr{\"o}ner's geometric theory of dislocation. 
Our gauge theoretical formulation of dislocation theory includes 
Kr{\"o}ner's basic equations and is thus a straightforward description
of dislocation theory with moment stress as given previously 
in~\cite{kroener69,trzesowski90}.

Kr{\"o}ner~\cite{kroener63}, Stojanovi{\'c}~\cite{stoja63}, and, later, 
Kleinert~\cite{kleinert83} have introduced the concept of a double gauge 
theory of dislocations, that is the stress tensor can be considered
as an Einstein tensor of a formal stress space with torsion.
From the geometrical and field theoretical point of view, there is no
need to interpret the stress as an Einstein tensor.
The generalization of the stress tensor is the energy-momentum tensor.
The stress tensor is nothing else but the source of the Einstein tensor,
and the equilibrium condition
\begin{align}
\text{div}\,{\sigma}=0
\end{align}
is the first Noether identity.
Therefore, we did not make use of the concepts of formal stress space and strain space.

Now some remarks are in place with respect to the moment stress caused by disclinations.
The generalization of (material) moment stress is the hypermomentum 
$\Delta_{ab}=\delta {\cal L}_{\mathrm{el}}/\delta\omega^{ab}$ which contains the 
spin current~\cite{gronwald}. The hypermomentum is the source of the 
$GL(n,{\Bbb R})$-gauge field in MAG and it does not appear in a pure translation gauge theory.
Therefore, there are no degrees of freedom for describing spin-disclination in 
a $T(3)$-gauge theory, in agreement with Kr{\"o}ner~\cite{disclination}. 
Spin-disclinations are defects in materials with
microstructure such as liquid crystals or magnetic spin systems,
where spin-moment stress occurs.
A gauge theory of materials with microstructure was given by
Lagoudas~\cite{lagoudas}.

The goal of this paper was the formulation of a dislocation theory 
as a gauge theory in analogy to gravity.  
We combined the physical ideas of Kr{\"o}ner's geometrical theory 
with the framework of MAG. 
We found a dislocation theory with moment stress $H_a$ and two kinds of force 
stresses $\Sigma_a$ and $h_a$.
In this picture, the dislocation is a source of nontrivial torsion in a 
Weitzenb{\"o}ck space or of Riemannian curvature in a Riemann space, respectively.
The elastic material plays the role of a kind of an (an)isotropic ``ether'' in 
analogy to the vacuum in gravity theory.
The metric (Cauchy-Green strain tensor) $g_{ij}$ is an effective quantity 
determined by $\phi^a$ and $\d\xi^a$. However, it is not a gauge potential.

\acknowledgments
The author is grateful to Profs. Friedrich W.~Hehl, Ekkehart Kr{\"o}ner, 
Luciano Mistura, Alfred Seeger, and Hans-Rainer Trebin and to Dr. Gerald Wagner for 
many helpful discussions, furthermore to Prof. Friedrich W.~Hehl for useful comments 
on an earlier version of this paper.
M.L. acknowledges the support by the Max-Planck-Institut f{\"u}r Metallforschung 
and the Institut f{\"u}r Theoretische \& Angewandte Physik,
University of Stuttgart  during his stay in Stuttgart.



\begin{references}
\bibitem{kondo52} K.~Kondo, {\it On the geometrical and physical foundations of the theory
	of yielding}, in {\it Proceedings of the 2nd Japan National Congress for Applied
	Mechanics}, Tokyo (1952) pp. 41-47
\bibitem{bilby55} B.A.~Bilby, R.~Bullough, and E.~Smith,
	Proc. Roy. Soc.~(London)~A~{\bf 231} (1955) 263
\bibitem{kroener59} E.~Kr{\"o}ner and A.~Seeger,
	Arch. Rat. Mech. Anal.~{\bf 3} (1959) 97 
\bibitem{kroener60} E.~Kr{\"o}ner, 
	Arch. Rat. Mech. Anal.~{\bf 4} (1960) 273 
\bibitem{kroener90} E.~Kr{\"o}ner, 
	Int. J. Theor. Phys.~{\bf 29} (1990) 1219 
\bibitem{kroener95} 
        E.~Kr{\"o}ner, {\it Dislocation theory as a physical field theory}, in:
        Continuum Models and Discrete Systems, Proc. 8th Int. Symp., Varna, Bulgaria, 
        K.Z. Markov ed., World Scientific, Singapore (1996) pp. 522-537
\bibitem{turski66} {\L}.~Turski, 
	Bull. Pol. Acad. Sci., S{\'e}r. techn. sci.~{\bf 14} (1966) 289 
\bibitem{edelen83} A.~Kadi{\'c} and D.G.B. Edelen, {\it A gauge theory of dislocations and disclinations},
        in {\it Lecture Notes in Physics}, Vol. 174, Springer, Berlin (1983)
\bibitem{edelen88} D.G.B.~Edelen and D.C.~Lagoudas, {\it Gauge theory and defects in 
        solids}, in {\it Mechanics and Physics of Discrete System}, Vol.~1,
	G.C.~Sih, ed., North-Holland, Amsterdam (1988)
\bibitem{hehl76} F.W.~Hehl, P. von der Heyde, G.D.~Kerlick, and J.~Nester,
	Rev. Mod. Phys.~{\bf 48} (1976) 393
\bibitem{gairola81} B.K.D.~Gairola,
	{\it Gauge invariant formulation of continuum theory of defects}, in:
 	Continuum Models and Discrete Systems, Proc. 4th Int. Symp.,  
        O.~Brulin and R.K.T.~Hsieh, eds., North-Holland, Amsterdam (1981) pp. 55-65
\bibitem{gairola93} B.K.D.~Gairola, {\it Gauge theory of dislocations}, in:
        Continuum Models and Discrete Systems, Proc. 7th Int. Symp., Paderborn, Germany, 
        K.-H.~Anthony and H.-J.~Wagner, eds.,
	Trans. Techn. Publ., Aedermannsdorf (CH) (1993) pp. 579-590
\bibitem{kleinert83} H.~Kleinert, 
	Phys. Lett.~A~{\bf 97} (1983) 51
\bibitem{kleinert89} H.~Kleinert, {\it Gauge fields in condensed matter Vol. II:
	Stresses and defects}, World Scientific, Singapore (1989) 
\bibitem{trzesowski93} A.~Trz\c{e}sowski, 
	Rep.~Math.~Phys.~{\bf 32} (1993) 71 
\bibitem{malyshev99} C.~Malyshev, {\it The dislocation stress functions from
	the double curl $T(3)$-gauge equation: Linearity and a look beyond}, 
	 cond-mat/9901316 (1999)
\bibitem{seeger55} A.~Seeger, {\it Theorie der Gitterfehlstellen}, in 
	{\it Handbuch der Physik VII/1}, S.~Fl{\"u}gge, ed., Springer, Berlin 
	(1955) pp. 383-665
\bibitem{hehl95} F.W.~Hehl, J.D. McCrea, E.W. Mielke, and Y. Ne'eman,
        Phys. Reports {\bf 258} (1995) 1
\bibitem{trautman80} A.~Trautman, 
        Gen.~Rel.~Grav.~{\bf 1} (1980) 287 
\bibitem{mielke} E.W.~Mielke, {\it Geometrodynamics of gauge fields},
        Akademie-Verlag, Berlin (1987)
\bibitem{koba1} S.~Kobayashi and K.~Nomizu, {\it Foundations of differential geometry}, 
        Vol.~1, Interscience Publishers, New York (1963)
\bibitem{trautman84} A.~Trautman, {\it Differential geometry for physicists}, 
        Stony Brook Lectures, Bibliopolis, Napoli (1984)
\bibitem{trautman73} A.~Trautman, {\it On the structure of the Einstein-Cartan equations}, 
        in {\it Differential Geometry, Symposia Mathematica Vol.~{\bf 12}}, 
        Academic Press, London (1973) pp. 139-162
\bibitem{mistura90} L.~Mistura, 
	Int. J. Theor. Phys.~{\bf 29} (1990) 1207 
\bibitem{schouten} J.A.~Schouten, {\it Ricci-Calculus}, 2nd ed., Springer, Berlin (1954)
\bibitem{kroener81} E.~Kr{\"o}ner, {\it Continuum theory of defects}, in
        {\it Physics of defects} (Les Houches, Session 35), R.~Balian et al., eds.,
        North-Holland, Amsterdam (1981) pp. 215-315
\bibitem{hehl98} F.W.~Hehl, personal communication (1998)
\bibitem{marsden} J.E.~Marsden and T.J.R.~Hughes, {\it Mathematical foundations of
        elasticity}, Dover, New York (1994)
\bibitem{Erice95} F.~Gronwald and F.W.~Hehl, {\it On the Gauge
  Aspects of Gravity}, in {\em Proc.\ of the 14th Course of the
    School of Cosmology and Gravitation on `Quantum Gravity',} held at
  Erice, Italy, May 1995, P.G.~Bergmann, V. de Sabbata, and H.-J.
  Treder, eds., World Scientific, Singapore (1996) pp. 148-198, gr-qc/9602013
\bibitem{MAG} F.~Gronwald, 
        Int. J. Mod. Phys.~D~{\bf 6} (1997) 263
\bibitem{thirring} W.~Thirring, {\it Classical Field Theory, A Course in Mathematical
	Physics~2}, 2nd ed., Springer, New York (1986)
\bibitem{cho} Y.M.~Cho,
	Phys. Rev.~D~{\bf 14} (1976) 2521
\bibitem{katanaev92} M.O.~Katanaev and I.V.~Volovich, 
	Ann.~Phys.~(NY)~{\bf 216} (1992) 1
\bibitem{hehl65} F.W.~Hehl and E.~Kr{\"o}ner, Z.~Naturforschg.~{\bf 20a} (1965) 336
\bibitem{osipov} V.A.~Osipov, J.~Phys.~A:~Math.~Gen.~{\bf 24} (1991) 3237
\bibitem{peach} M.~Peach and J.S.~Koehler, 
	Phys. Rev.~{\bf 80} (1950) 436
\bibitem{electro1} R.A.~Puntigam, C.~L{\"a}mmerzahl, and F.W.~Hehl,
	Class. Quant. Grav.~{\bf 14} (1997) 1347 
\bibitem{electro2} F.W.~Hehl, Y.N.~Obukhov, and G.F.~Rubilar, {\it Classical
	electrodynamics: A tutorial on its foundations}, in
	{\it Quo vadis geodesia...?} Festschrift for Erik W.~Grafarend, 
	F.~Krumm and V.S.~Schwarz, eds., Univ. Stuttgart, ISSN 0933-2839 (1999)
	pp. 171-184, physics/9907046
\bibitem{kroener69} E.~Kr{\"o}ner, {\it Plastizit{\"a}t und Versetzungen}, in
        {\it Mechanik der deformierten Medien} by A.~Sommerfeld, Harri
	Deutsch, Frankfurt/M. (1992) reprint of the 6th ed. (1969) 
\bibitem{trzesowski90} A.~Trz\c{e}sowski and J.J.~S{\l}awianowski, 
	Int. J. Theor. Phys.~{\bf 29} (1990) 1239 
\bibitem{kroener63} E.~Kr{\"o}ner, 
        Ann. Phys.~(Leipzig)~{\bf 11} (1963) 13 
\bibitem{stoja63} R.~Stojanovi{\'c}, 
	Int. J. Engng. Sci.~{\bf 1} (1963) 323 
\bibitem{gronwald} F.~Gronwald and F.W.~Hehl, {\it Stress and hyperstress as fundamental
        concepts in continuum mechanics and in relativistic field theory},
        in {\it Advances in Modern Continuum Dynamics}, International Conference 
	in Memory of Antonio Signorini, Isola d'Elba, June 1991, G. Ferrarese, 
	ed. Pitagora Editrice, Bologna (1993) pp. 1-32, gr-qc/9701054
\bibitem{disclination} E.~Kr{\"o}ner and D.C.~Lagoudas, 
        Int. J. Engng. Sci.~{\bf 30} (1992) 47 
\bibitem{lagoudas} D.C.~Lagoudas, 
        Int. J. Engng. Sci.~{\bf 27} (1989) 237 
\end{references}
\end{document}